# Protein Fold Family Recognition From Unassigned Residual Dipolar Coupling Data

Rishi Mukhopadhyay[1], Paul Shealy[1], and Homayoun Valafar[1],
[1]Department of Computer Science and Engineering, University of South Carolina, Columbia, SC, USA

**Abstract** – *Despite many advances in computational modeling of protein structures, these methods have not been widely utilized by experimental structural biologists. Two major obstacles are preventing the transition from a purely-experimental to a purely-computational mode of protein structure determination. The first problem is that most computational methods need a large library of computed structures that span a large variety of protein fold families, while structural genomics initiatives have slowed in their ability to provide novel protein folds in recent years. The second problem is an unwillingness to trust computational models that have no experimental backing. In this paper we test a potential solution to these problems that we have called Probability Density Profile Analysis (PDPA) that utilizes unassigned residual dipolar coupling data that are relatively cheap to acquire from NMR experiments.*

## 1   Introduction

Even as techniques for the computational modeling of protein structures from amino acid sequences have become increasingly sophisticated, the structures produced by these methods are still hampered by two critical hurdles. First, most of the successful methods [1-4] require extensive databases of experimentally derived structures that represent a wide variety of protein fold families. The growth of these databases, however, has slowed tremendously in the past few years. Based on the statistics provided by the PDB [5], using CATH [6] classifications of fold families indicates that only two novel families were discovered in the year 2005 and no additional ones since then. Current methods of novel target selection by structural genomics initiatives, therefore, appear to be in need of an overhaul.

The second obstacle faced by computational structure determination algorithms is that the results from even the most well-tested methods generally aren't trusted. Most such methods, like [1-4] usually return a list of candidate structures. While the actual structure may be close to one of the candidates returned by these programs, there is no generally reliable method of ranking the candidates. Even if there were, it is unlikely that the results would be considered sufficiently trustworthy by the community of experimental biologists. Therefore, there exists a need for a technique that could use a small amount of experimental data to rank and/or validate candidate structures produced by computational methods.

Previous research [7] has begun to address both of these issues within a limited scope. The previous work demonstrated the possibility of identifying the structure 1C99 among a library of 21 structures using simulated RDCs from the backbone N-H vectors. This library of 21 structures represented 9 different FSSP protein fold families. RDC data can be acquired rapidly, inexpensively and accurately by nuclear magnetic resonance (NMR) spectroscopy. The result of this investigation is that in the absence of noise, PDPA was confirmed as a technique that could use unassigned residual dipolar coupling data to correctly rank the other members of a protein's fold family as its closest relatives from a database of candidates. This study was then extended to the case of noisy data, and even then PDPA was able to identify the other members of 1C99's fold family as its nearest relatives.

In this paper we seek to extend this method to a search of all 619 protein fold family representatives in FSSP (as of 2003) to see if PDPA can be used for fast and cheap protein fold family identification. Additionally, we will introduce a refinement of this method that incorporates the use of correlated $C^\alpha$-$H^\alpha$ Residual Dipolar Couplings.

## 2   Methods and Methods

### 2.1   Residual Dipolar Couplings

Residual dipolar couplings arise from the interaction between spin 1/2 nuclei (e.g.: $^{15}$N-$^{1}$H) in strong magnetic fields according to equation 1 where the angle brackets denote an averaging over time and theta is the angle between the interatomic vector and the magnetic field.

$$D_{ij}^{res} = -\left(\frac{\mu_0}{4\pi}\right)\frac{\gamma_i \gamma_j h}{2\pi^2 r_{ij,eff}^3} \langle \frac{3 \cdot \cos^2\theta - 1}{2} \rangle \quad (1)$$

For isotropically tumbling molecules, the probability density function is shown in equation 2.

$$p(\theta) = \sin(\theta)/2 \quad (2)$$

The final observable RDC values will be reduced to zero for an isotropically tumbling molecule as shown in equation 3.

$$\int_0^\pi \frac{(3 \cdot \cos^2\theta - 1)}{2} \sin\theta \cdot \partial\theta = 0 \qquad (3)$$

Since this integral evaluates to 0, residual dipolar coupling phenomena are not generally observed. However, anisotropy, or partial alignment, can be induced by dissolving a sample in a solution that contains an aligning solute such as bicelle or filamentous phage [8]. Saupe observed, that rather than having to know the exact distribution of the molecule's tumbling to make calculations about RDCs, the nature of the quantity being time-averaged allows the molecule's tumbling to be described by a 5-element order tensor. For Cartesian coordinates, the equation for the time-averaged RDC of a normalized vector can be described as shown in equation 4,

$$D = v^T \cdot \begin{bmatrix} s_{xx} & s_{xy} & s_{xz} \\ s_{xy} & s_{yy} & s_{yz} \\ s_{xz} & s_{yz} & s_{zz} \end{bmatrix} \cdot v \qquad (4)$$

where $s_{xx}$, $s_{syy}$, $s_{zz}$, $s_{xy}$, $s_{xz}$, $s_{yz}$ are the variables describing the alignment of the molecule in the aligning medium, and $s_{xx} + s_{yy} + s_{zz} = 0$. This order tensor can then be decomposed by the use of singular value decomposition into $RSR^T$ where $R$ is a rotation matrix and $S$ is a diagonal matrix. This form can be conceptualized as rotating the molecule into a principal alignment frame and then using the simpler equation 5 to calculate the RDCs.

$$D = x^2 \cdot s_{xx} + y^2 \cdot s_{yy} + z^2 \cdot s_{zz} \qquad (5)$$

The rotation matrix is then decomposed into three Euler rotations: $R = R_z(\alpha) \cdot R_y(\beta) \cdot R_z(\gamma)$. In this formulation, the five variables used to describe the anisotropic tumbling of the molecule are $\alpha$, $\beta$, $\gamma$, $s_{xx}$, [9] and $s_{yy}$.

## 2.2 1D-PDPA

We first introduce the concepts of "query" and "subject" proteins in order to facilitate further discussions. A query protein is the protein for which experimental data has been acquired and structural information is sought. A subject protein is the protein for which a detailed atomic description of structure already exists in a library of structures. A PDP for a set of RDCs is defined as the application of a Parzen density operator to a set of RDC's with a bandwidth chosen to reflect the expected experimental error. The PDP of a query protein can be obtained using experimental data (denoted as *ePDP*). The PDP of a subject protein can be obtained using RDCs computed from the structure of the protein and a given order tensor (denoted as *cPDP*). A comparison of *ePDP* and *cPDP* can provide a measure of structural similarity between the query and subject proteins. This process of utilizing PDPs to obtain structural similarity between two proteins is referred to as Probability Density Profile Analysis (*PDPA*).

A number of impediments rooted in innate properties of RDC data stand in the way of simply comparing two PDPs in order to ascertain structural homology. First, PDPs depend on orientation of protein structures; that is, two identical structures can produce completely different PDPs when aligned differently with respect to the external magnetic field $B_0$. Second, it is possible for two completely different structures to produce identical PDPs if the structural relationship between the two structures perfectly coincides with symmetric properties of the alignment tensor, such as inversion. The stated problem exhibits a superficial similarity to identification of objects based on their shadows. Two different objects such as a cylinder and a cone may produce similar circular shadows under some coincidental orientation of the two objects. On the other hand, two identical cylindrical objects may produce two completely different shadows (such as circular and rectangular) depending on their orientations in the field of illumination. The second impediment can be resolved by an exhaustive exploration of all possible orientations of the subject protein. This way, any two similar structures should produce at least one instance of similar PDPs at some alignment of the subject protein. The first impediment is simply rooted in symmetric properties of RDC data; in the second part of this investigation, we propose to address this potential problem by using RDC data from $C^\alpha$-$H^\alpha$ vectors in addition to the N-H vectors. While it is unlikely that two structures may share a structural relationship that perfectly coincidences with the symmetrical properties of the first alignment medium, occurrence of this phenomenon for two types of vectors should be highly unlikely if not impossible.

In the limit, a large protein with randomly distributed N-H vectors will have a PDP that converges to the relatively featureless powder pattern shown in Figure 1. However, in practice, proteins suitable for NMR are smaller in size and secondary structural elements such as beta sheets and alpha helices impose order on the distribution of the orientation of N-H vectors. In Figure 1, the PDP of protein ARF (1HUR) is shown superimposed on the corresponding powder pattern. In essence it is this deviation from the powder pattern that forms the structural fingerprint that is utilized by the PDP.

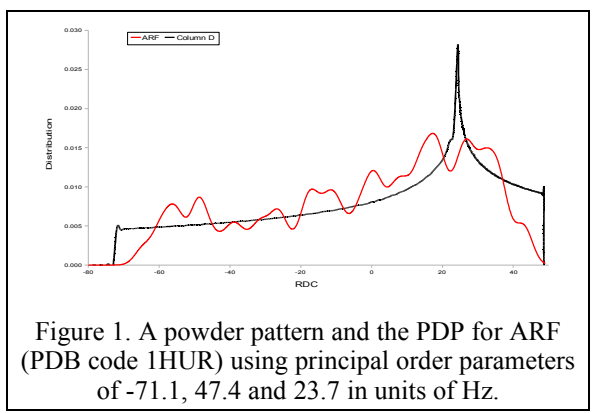

Figure 1. A powder pattern and the PDP for ARF (PDB code 1HUR) using principal order parameters of -71.1, 47.4 and 23.7 in units of Hz.

In general, a PDP of any given structure depends on three components: its tertiary structure, principal order parameters and orientational alignment of the protein. Therefore, a thorough approach to comparing query and subject proteins is the construction of an algorithm that conducts a search over all structures, order parameters, and possible orientations of each structure. This approach would require an impractical computation time. However, the search over the entire space of principal order parameters can be confined by estimation of order parameters from the experimentally observed PDP (or the ePDP). The attainment of the principal order parameters from an unassigned list of RDC data has been previously demonstrated [7; 9-11]. The search over all protein structures (currently over 30,000 structures in PDB) can be limited to representatives of each fold family and therefore be reduced to the 1056 protein fold family representatives reported by SCOP [12] as of March, 2008. The remaining search of all orientations can be achieved by simply altering the presumed alignment of any given protein without any change in the principal order parameters. The resolution of the grid search is selected based on available computational resources. For this study, a resolution of 5° was chosen.

Selection of an appropriate metric in quantifying the similarity of two PDP maps is very critical. We have considered a large number of different metrics, such as correlation coefficient, root-mean-squared-deviation (rmsd), and Euclidean distance, which have been used successfully in other fields [13; 14]. Based on this consideration, we have selected a modified $\chi^2$ scoring scheme for our studies. The conventional $\chi^2$ score is not appropriate, because it does not produce a symmetric report of the distance between two patterns; that is, for patterns A and B, $\chi^2(A,B) \neq \chi^2(B,A)$. The main goal of our modification is to eliminate this lack of symmetry while reducing the harsh penalty of missing data. Equations 6 and 7 define the scoring mechanism used in this research. $S(cPDP, ePDP)$ in Equation 6 denotes the final comparison score between cPDP and ePDP. The summation index M denotes the number of points that are sampled in comparing the two PDPs. Entities $c_i$ and $e_i$ indicate the values of computed and experimentally determined PDPs at the location $i$, respectively. The distance at any given position of two PDPs is determined by $\chi^2(c,e)$ as defined in Equation 7.

$$S(cPDP, ePDP) = \frac{1}{2} \sum_{i=1}^{M} \left[ \chi^2(c_i, e_i) + \chi^2(e_i, c_i) \right] \quad (6)$$

$$\chi^2(c,e) = \begin{cases} \dfrac{(c-e)^2}{c} , & c \neq 0 \\ 100\,e , & c = 0 \end{cases} \quad (7)$$

In [7], PDPA was used to successfully identify 1C99 from a list of 21 candidate structures representing 9 protein fold families. Furthermore, it was able identify the other members of 1C99's fold family as its nearest relatives. PDPA was also shown to be able to continue this success, even in the presence of noise.

### 2.3 2D-PDPA

While identification of an unknown protein among the library of fold families (~1000 proteins) is of tremendous use, it would be of great interest to explore the possibility of identifying the most homologous structure from the entire PDB database (~40,000 proteins). This significant expansion in the scope of the *PDPA* is likely to require additional information.

In principle, *PDPA* does not have to be confined to only one set or type of data. Upon the identification of an appropriate alignment medium for a given protein, acquisition of additional RDC data (such as $C_\alpha$-$H_\alpha$) or perturbation of the alignment will have very little impact on the total acquisition time. A minimal extension of data acquisition period can provide significantly more experimental data such as two sets of RDC data (from N-H and $C_\alpha$-$H_\alpha$ or just the N-H data from two different media). Integration of additional data is anticipated to substantially increase the information content and therefore significantly improve the robustness and sensitivity of *1D-PDPA*. Therefore, the second part of this paper, focuses on the extension of the above method to incorporate data from $C_\alpha$-$H_\alpha$ vectors in addition to N-H vectors for simultaneous analysis. The additional data can be incorporated without the need for assignment since chemical shift data can be used to pair RDC values originated from the same residue. Just like the above procedure, this *2D-PDPA* procedure will use parzen density estimation to produce PDP's and use the obvious extension of equations 1 and 2 to compute the distance between the *ePDP* and each *cPDP*. Figure 2 illustrates 2D-PDPs for two different proteins: *Galectin-3* and *Myoglobin* (PDB codes of 1A3K and 1I0M respectively) using simulated backbone N-H and $C_\alpha$-$H_\alpha$ RDC data. These data have been generated with an arbitrary alignment tensor and addition of appropriate noise (2 Hz for N-H and 4 Hz for $C_\alpha$-$H_\alpha$). The degree of dispersion of peaks in transition from 1D to 2D *PDPA* shares a great deal of conceptual parallels with 1D and 2D spectroscopy. Overlapping peaks in 1D are distinctly resolved in 2D when observed in the presence of additional data.

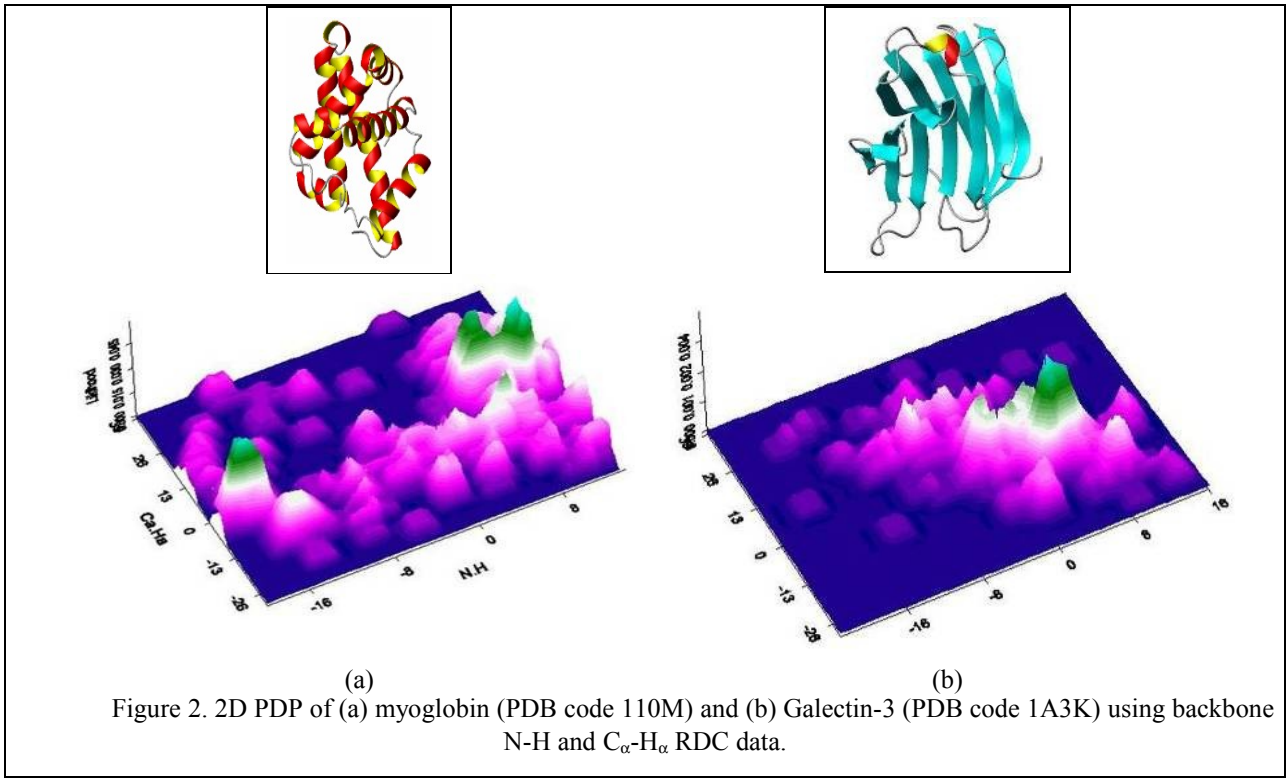

(a) (b)

Figure 2. 2D PDP of (a) myoglobin (PDB code 1I0M) and (b) Galectin-3 (PDB code 1A3K) using backbone N-H and $C_\alpha$-$H_\alpha$ RDC data.

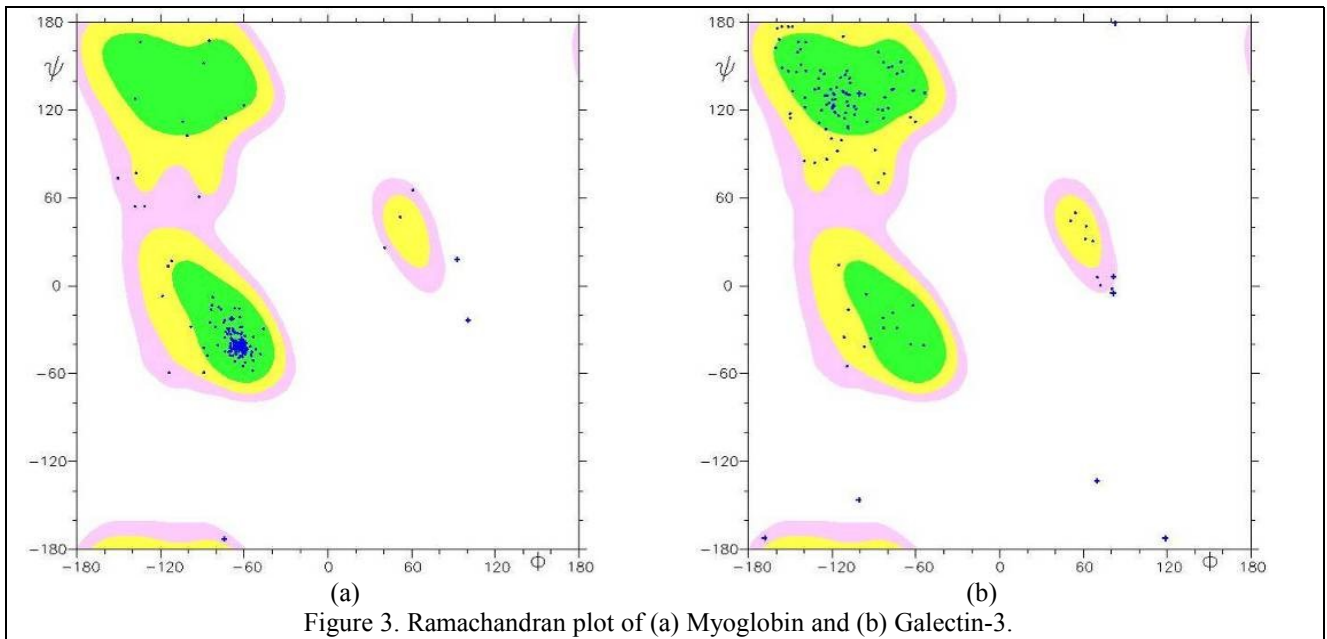

(a) (b)

Figure 3. Ramachandran plot of (a) Myoglobin and (b) Galectin-3.

A *2D-PDPA* is expected to improve the capabilities of *1D-PDPA* in a number of ways. First, dispersing structurally characteristic peaks in two dimensions improve specificity and robustness of PDPA. Second, sampling of RDC space provided by two sets of data is more likely to provide a better estimate of the order parameters. Accurate estimation of the order parameters is clearly an important factor in the performance of the PDPA.

## 3 Results and Discussion

### 3.1 *1D-PDPA* using backbone N-H RDCs

The first results obtained from *1D-PDPA* aims to establish the feasibility of deploying this methodology in identification of the correct structure within a library of fold families (~619 structures). The protein 16VPA was randomly selected as the unknown structure and experimental data was simulated for this protein with 1 Hz error and an arbitrary but typical order tensor. Each of the

619 FSSP (as of 2003) protein fold family representatives was then compared to the *ePDP* of 16VPA by using the estimated principal order parameters and calculating a *cPDP* for each protein for all possible orientations within 5° resolution. The best match (lowest modified chi-squared score) was then used as that protein's similarity score. Figure 4 shows a graph of the scores generated in this fashion. Ideally, the actual structure of 16VPA would have been the best $\chi^2$ match amongst all the candidates. Although 16VPA ranked 8th, certain additional information can be used to improve this ranking. Assuming that proteins that are significantly different in size are unlikely to exhibit significant structural homology, half of the proteins that ranked as better matches to 16VPA than itself could have been eliminated on the basis of being more than twice as large. This result indicates a need for improvement through the addition of more data, however.

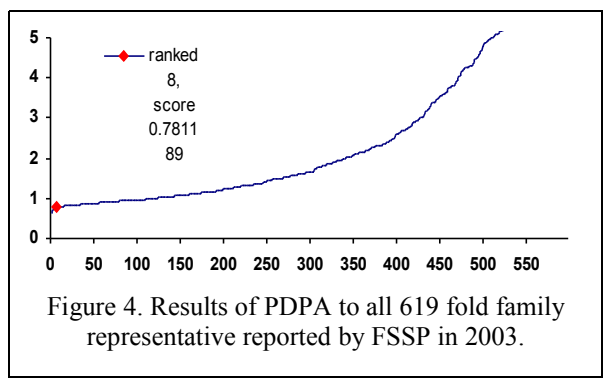

Figure 4. Results of PDPA to all 619 fold family representative reported by FSSP in 2003.

### 3.2 *2D-PDPA* on N-H and $C^α$-$H^α$ RDCs

In addition to the RDCs for N-H vectors, $C_α$-$H_α$ RDCs can be collected and paired with the N-H vectors from the same peptide planes. This data has the potential to reveal a lot about the structure of a protein. For instance, φ angles of approximately -120 (or 60) creates the anti-parallel (or parallel) arrangement of the backbone N-H or $C_α$-$H_α$ that leads to secondary structural elements like beta-sheets. When this is the case, the RDC values for N-H and $C_α$-$H_α$ vectors will be the same once they are corrected for different gyromagnetic ratios and bond distances, and therefore occupy the diagonal portion of the 2D-PDP.

Simple observation of the fraction of peaks located in the diagonal region of a *2D-PDP* will provide the fraction of φ torsion bonds within the vicinity of -120° (or 60°). Figure 2 (a) represents the 2D-PDP corresponding to Myoglobin (PDB code 1IOM) and Figure 2 (b) represents the *2D-PDP* corresponding to human Galectin-3 (PDB code 1A3K). Figure 3 shows the Ramachandran plots of the two structures. Note the correlation between the convergence of the peaks in the 2D-PDP map and clustering of the angular information that clusters within the α-helical region of the Ramchandran space.

Applying the same modified $\chi^2$ score to the 2D-PDP, corresponding to the N-H- $C_α$-$H_α$ distributions to the top 8 candidates from *1D-PDPA* analysis correctly placed 16VPA at the top of the list as the best candidate as shown in Table 1. While this experiment is relatively small-scale, combined with the 1D-PDPA, it demonstrates the usefulness of incorporating additional correlated channels of data to improve the discriminatory power of *PDPA*.

Table 1. Results of 2D-PDPA to top 8 structures selected by 1D-PDPA.

| PDB ID | *2D-PDPA* Score |
|---|---|
| 16vpA, | 761.17 |
| 1qi7A, | 793.18 |
| 1nzyA, | 834.81 |
| 1oasA, | 965.5 |
| 1qlmA, | 976.29 |
| 1dd8A, | 1226.8 |
| 1f0xA, | 1410.6 |
| 1ddzA, | 1676.4 |

## 4 Conclusion

The preliminary trial of the *PDPA* system for identifying protein fold family from a large set of protein fold family representatives demonstrated a high degree of success and, most importantly, indicates that the addition of extra channels of data can greatly enhance the discriminatory power of *PDPA*. Consequently, it can have a significant impact in a number of areas of global research. Existence of a continuum of tools that can potentially emerge from our research will facilitate a natural path of transition toward the desirable and anticipated *in silico* biological studies. A number of related and parallel fields of endeavor will also take advantage of these findings. For example, large sums of resources have already been allocated toward the expeditious completion of the protein fold space. A number of other critical areas of research such as protein threading rely heavily on the completion of this database. Analysis of the recent depositions to the protein data bank (PDB, www.rcsb.org) reveals a 10% success in identification of novel structures by the general community of structural biologist and less than 20% (data published on individual sites) by the structural genomics centers dedicated to this task . This 80-90% inefficiency on the part of the structural genomics centers in targeting proteins of interest will have a great impact in completion of the comprehensive library of protein folds. At this rate, the library of 10,000 structures can be extrapolated to take at least an additional 20-30 years. Completion of this research can lead to development of more effective protein screening tools with significantly better performance in isolating structurally novel proteins.

Currently, the performance of threading algorithms is heavily tied to sequence homology. Information produced from this research can lead to accurate fold classification of an unknown protein. This information can then be used to provide a structural template, independent of sequence homology. The current 30% sequence homology barrier can then be breached. The utility of such a tool is not only limited to determination of appropriate method of structure determination; it can also be used to select the most appropriate structural template based on structural information and not necessarily sequence homology. Furthermore, it can serve as an independent method of scrutinizing obtained structures whether by X-Ray crystallography, conventional distance based NMR spectroscopy, or computational methods. Within the community of investigators, concerns have been expressed about the diminishment of structural quality when produced under the constraints of rapid and high-throughput methods.

### 4.1 Future Direction

Albeit PDPA has demonstrated a great success in identifying homologous structures with a limited set of data, it is easy to argue that there exists room for further improvements. Among several improvements, the proper construction and scoring mechanism between two PDP patterns remains to be one of the most critical needs for integration of this analysis in large-scale applications.

The second important problem that is known to reduce the efficacy of PDPA is the general influence of powder pattern on individual PDPs. The information content of PDPs is attenuated nonlinearly by the general shape of a powder pattern, especially at the large middle peak of a powder pattern (highest peak in ). Assessment of likelihoods needs to reflect the structural influence without the general influence of powder pattern that is primarily dictated by the three principal order parameters. Elimination of this problem is anticipated to have a significant contribution in both sensitivity and selectivity of future versions of PDPA.

Through the development of this methodology we hope to be able to provide a highly adaptable approach that can be utilized under a spectrum of scenarios. Such as tools will be able to provide a template structure from minimum set of data and an extensive library of structures at one extreme of the spectrum. At the other end of the spectrum, we endeavor to extend this tool to provide a significant amount of structural information in the presence of large amount of data and a very small library of structures.

## 5 Acknowledgement

Funding for this project was provided by NSF Career Grant # MCB-0644195

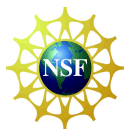